\documentclass[conference]{IEEEtran}
\IEEEoverridecommandlockouts
\usepackage{cite}
\usepackage{amsmath,amssymb,amsfonts}
\usepackage{algorithmic}
\usepackage{graphicx}
\usepackage{textcomp}
\usepackage{xcolor}
\usepackage{physics}
\usepackage{quantikz}
\usepackage{booktabs}
\usepackage{verbatim}
\usepackage{siunitx}
\usepackage[colorlinks=true, linkcolor=blue, citecolor=blue, urlcolor=blue]{hyperref}
\usepackage{float}
\usepackage{stfloats}
\usepackage{booktabs}
\usepackage{pifont}      
\newcommand{\cmark}{\ding{51}}  
\newcommand{\xmark}{\ding{55}}  
\usepackage{booktabs,pifont}

\usepackage{algorithm}
\usepackage{algorithmic}
\def\BibTeX{{\rm B\kern-.05em{\sc i\kern-.025em b}\kern-.08em
    T\kern-.1667em\lower.7ex\hbox{E}\kern-.125emX}}

\newcommand{\Call}[2]{\textsc{#1}(#2)}

\begin{document}

\title{Low-Level and NUMA-Aware Optimization for High-Performance Quantum Simulation\\
}

\author{\IEEEauthorblockN{1\textsuperscript{st} Ali Rezaei}
\IEEEauthorblockA{{The University of Edinburgh} \\
Edinburgh, UK \\
ali.rezaei@ed.ac.uk}
\and
\IEEEauthorblockN{2\textsuperscript{nd} Luc Jaulmes}
\IEEEauthorblockA{{The University of Edinburgh} \\
Edinburgh, UK \\
ljaulmes@exseed.ed.ac.uk}
\and
\IEEEauthorblockN{3\textsuperscript{rd} Maria Bahna}
\IEEEauthorblockA{{The University of Edinburgh} \\
Edinburgh, UK \\
maria.bahna@ed.ac.uk}
\and
\IEEEauthorblockN{4\textsuperscript{th} Oliver Thomson Brown}
\IEEEauthorblockA{{EPCC} \\
Edinburgh, UK \\
o.brown@epcc.ed.ac.uk}
\and
\IEEEauthorblockN{5\textsuperscript{th} Antonio Barbalace}
\IEEEauthorblockA{{The University of Edinburgh} \\
Edinburgh, UK \\
antonio.barbalace@ed.ac.uk}}

\maketitle

\begin{abstract}
Scalable classical simulation of quantum circuits is crucial for advancing quantum algorithm development and validating emerging hardware.
This work focuses on performance enhancements through targeted low-level and NUMA-aware tuning on a single-node system, thereby not only advancing the efficiency of classical quantum simulations but also establishing a foundation for scalable, heterogeneous implementations that bridge toward noiseless quantum computing. 
Although few prior studies have reported similar hardware-level optimizations, such implementations have not been released as open-source software, limiting independent validation and further development.
We introduce an open-source, high-performance extension to the QuEST state vector simulator that integrates state-of-the-art low-level and NUMA-aware optimizations for modern processors. 
Our approach emphasizes locality-aware computation and incorporates hardware-specific techniques including NUMA-aware memory allocation, thread pinning, AVX-512 vectorization, aggressive loop unrolling, and explicit memory prefetching. 
Experiments demonstrate substantial speedups--5.5-6.5$\times$ for single-qubit gate operations, 4.5$\times$ for two-qubit gates, 4$\times$ for Random Quantum Circuits (RQC), and 1.8$\times$ for the Quantum Fourier Transform (QFT). 
Algorithmic workloads further achieve 4.3--4.6$\times$ acceleration for Grover and 2.5$\times$ for Shor-like circuits. These results show that systematic, architecture-aware tuning can significantly extend the practical simulation capacity of classical quantum simulators on current hardware.
\end{abstract}

\maketitle

\section{Introduction}
Classical simulation of quantum circuits remains invaluable for validating algorithms, benchmarking hardware, and debugging control software while error-corrected quantum computers remain beyond current reach. In the full state‑vector simulation an $n$-qubit state is represented by $2^n$ complex amplitudes, so both the memory footprint (16 bytes per amplitude in double-precision) and the number of floating‑point operations grow exponentially with $n$. Even a modest 35‑qubit circuit already requires about 0.5 TiB of memory and billions of complex arithmetic operations, quickly exhausting today’s servers. Scaling beyond a few dozen qubits remains challenging because hardware and software bottlenecks dominate performance. Memory bandwidth and capacity are primary constraints: even on multi‑socket non‑uniform memory access (NUMA) servers, accessing a $2^n$‑element state vector saturates the memory hierarchy~\cite{Ref1}. In distributed quantum-circuit simulation, the network communication required for qubit gates--whether single- or two-qubit--spanning different nodes introduces substantial overhead and often dominates the run time~\cite{Ref1,Ref2}. At the software level, most simulators still under-use low‑level hardware features, leaving appreciable performance untapped and underscoring the need for architecture‑aware refinement.

\textbf{\textit{State-of-the-art and practice.}}
To push the simulation limits, several high‑performance state‑vector simulators have emerged that exploit modern hardware architectures and parallelism~\cite{Ref3}. QuEST~\cite{Ref4}, ProjectQ~\cite{Ref5}, Intel‑QS (qHiPSTER)~\cite{Ref6}, Qrack~\cite{Ref7}, Qulacs~\cite{Ref8}, Quantum++~\cite{Ref9}, qsim~\cite{Ref10}, and more recently PennyLane Lightning~\cite{Ref11} are representative state-of-the-art examples -- summarized in Table~\ref{tab:sim-optimisations}. 
Among many simulators exploiting multi‑threading (OpenMP), distributed Message Passing Interface (MPI) execution, or GPU off‑loading to enlarge the range of tractable problem sizes, some state vector simulators accelerate the core operation--applying gate matrices to the state vector--through explicit single-instruction-multiple-data (SIMD) parallelism. 
By operating on multiple amplitudes per CPU instruction, SIMD vectorization can significantly speed up state updates~\cite{Ref4}. 
For example, the ProjectQ simulator explicitly supports Intel AVX vector instructions \cite{Ref5}. 
More recently, Google’s qsim is built with AVX2 and fused multiply-add (FMA) intrinsics, using fused gate kernels to exploit data-level parallelism \cite{Ref10}. 
Xanadu’s Lightning similarly employs explicit AVX-512 intrinsics, effectively unrolling loops to operate on 8 or 16 values at a time \cite{Ref11}. 
These approaches reduce instruction overhead by applying gate operations to multiple state vector entries concurrently. 
Academic simulators have reported substantial speedups from SIMD optimizations. 
The HpQC simulator used AVX2/FMA instructions to maximize SIMD utilization, combined with manual bit-level optimizations \cite{Ref12}, yielding an average 2.20x speedup (GNU compiler) over QuEST on QFT benchmarks using a multi-core CPU. The authors attribute this gain largely to vectorized gate kernels and reduced instruction count. Another effort, PAS \cite{Ref13}, introduced a hybrid vectorization scheme alongside other techniques, achieving multi-fold speedups versus QuEST.
\begin{table*}[t]
\centering
\caption{Hardware-oriented optimisations in leading CPU-centric quantum simulators.}
\label{tab:sim-optimisations}
\small
\setlength{\tabcolsep}{3pt}
\begin{tabular}{@{}lcccccccccc@{}}
\toprule
\textbf{Simulator} &
\textbf{Vector\-isation} &
\textbf{Manual\,Unroll} &
\textbf{Cache\,Blocking} &
\textbf{OMP} &
\textbf{Thread\,Pinning} &
\textbf{NUMA} &
\textbf{MPI} &
\textbf{GPU} &
\textbf{Open-Source} \\ \midrule
ProjectQ        & AVX2 (comp.)       & \xmark & \cmark & \cmark & ---            & \xmark & \xmark & \xmark & \cmark \\
qsim            & AVX2 (intrin.)     & \xmark & \cmark & \cmark & ---            & \xmark & \xmark & \xmark & \cmark \\
Lightning       & AVX2/512 (intrin.) & \cmark & \xmark & \cmark & \textit{likely} & \xmark & \cmark & \cmark & \cmark \\
HpQC/PAS        & AVX2 (intrin.)     & \xmark & \xmark & \cmark & ---            & \xmark & \cmark & \xmark & \xmark \\
Qulacs          & AVX2 (intrin.)     & \xmark & \xmark & \cmark & ---            & \xmark & \xmark & \cmark & \textit{\cmark} \\
Qiskit Aer      & AVX2 (comp.)       & \xmark & \xmark & \cmark & ---            & \xmark & \xmark & \cmark & \cmark \\
Intel-QS        & AVX512 (comp.)     & \xmark & \xmark & \cmark & \textit{likely} & \xmark & \cmark & \xmark & \cmark \\
QuEST           & AVX2 (comp.)       & \xmark & \xmark & \cmark & ---            & \xmark & \cmark & \cmark & \cmark \\
NWQ-Sim         & AVX2 (comp.)       & \xmark & \xmark & \cmark & ---            & \xmark & \cmark & \cmark & \cmark \\
\bottomrule
\end{tabular}

\vspace{2pt}
\footnotesize
“Aligned Memory”, not included, is largely implicit since most simulators use 64-byte alignment for AVX-512 operations.  
“Thread Pinning” and “NUMA-Aware Allocation” are seldom addressed explicitly; in most cases, thread affinity follows the default OpenMP scheduler unless the user enforces binding via environment variables such as \texttt{OMP\_PROC\_BIND} or \texttt{OMP\_PLACES}.  
PennyLane Lightning and Intel-QS are inferred to support such affinity through OpenMP configuration rather than hard-coded pinning.  
\emph{comp.}~= compiler auto-vectorisation; \emph{intrin.}~= handwritten intrinsics.  
\emph{likely}~= affinity supported or recommended (e.g., via OpenMP settings) but not explicitly enforced in source code.  
“---”~= no explicit thread-affinity or pinning support (default OpenMP scheduling).  
\cmark/\xmark~= feature explicitly present or absent according to publication, documentation, or verified source inspection.
\end{table*}

To fully leverage SIMD, simulators ensure proper memory alignment of the state vector (often 32- or 64-byte aligned for AVX2/512). 
Although often not explicitly discussed in publications, memory alignment is a standard practice to avoid penalties when loading vectors into registers. 
Many simulators also focus on cache-friendly memory access patterns. A common technique is cache blocking--structuring computations to maximize reuse of data in fast caches before accessing slower memory.
In quantum-circuit terms, this translates to fusing consecutive gate operations so that the state vector is traversed fewer times. 
At a higher level, circuits can also be rearranged to minimize communication, effectively cache-blocking at the node level \cite{Ref14,Ref15}. ProjectQ and Google’s qsim use such gate fusion to improve locality. By multiplying small gate matrices into a larger effective matrix, the simulator can apply a batch of operations with one pass over the state, thus improving cache utilization and reducing memory bandwidth pressure. Beyond fusion, recent research introduced novel data layouts and filters to minimize unnecessary memory traffic. PAS describes a “memory access filtering” method, which avoids touching state vector regions that are unaffected by a given gate. Likewise, PAS’s use of fast bitwise operations optimizes the index calculations for mapping qubit operations to memory locations, replacing expensive arithmetic or branch-heavy code with efficient bit-manipulation. The HpQC tool also emphasized a custom data structure to exploit spatial locality in the cache. In practice, these low-level memory optimizations yielded measurable speedups: e.g., HpQC reported 1.74$\times$ (GNU compiler) speed improvement on RQCs vs. baseline QuEST by cutting down redundant memory accesses.
\begin{figure}[!htbp]
\centerline{\includegraphics[width=0.5\textwidth]{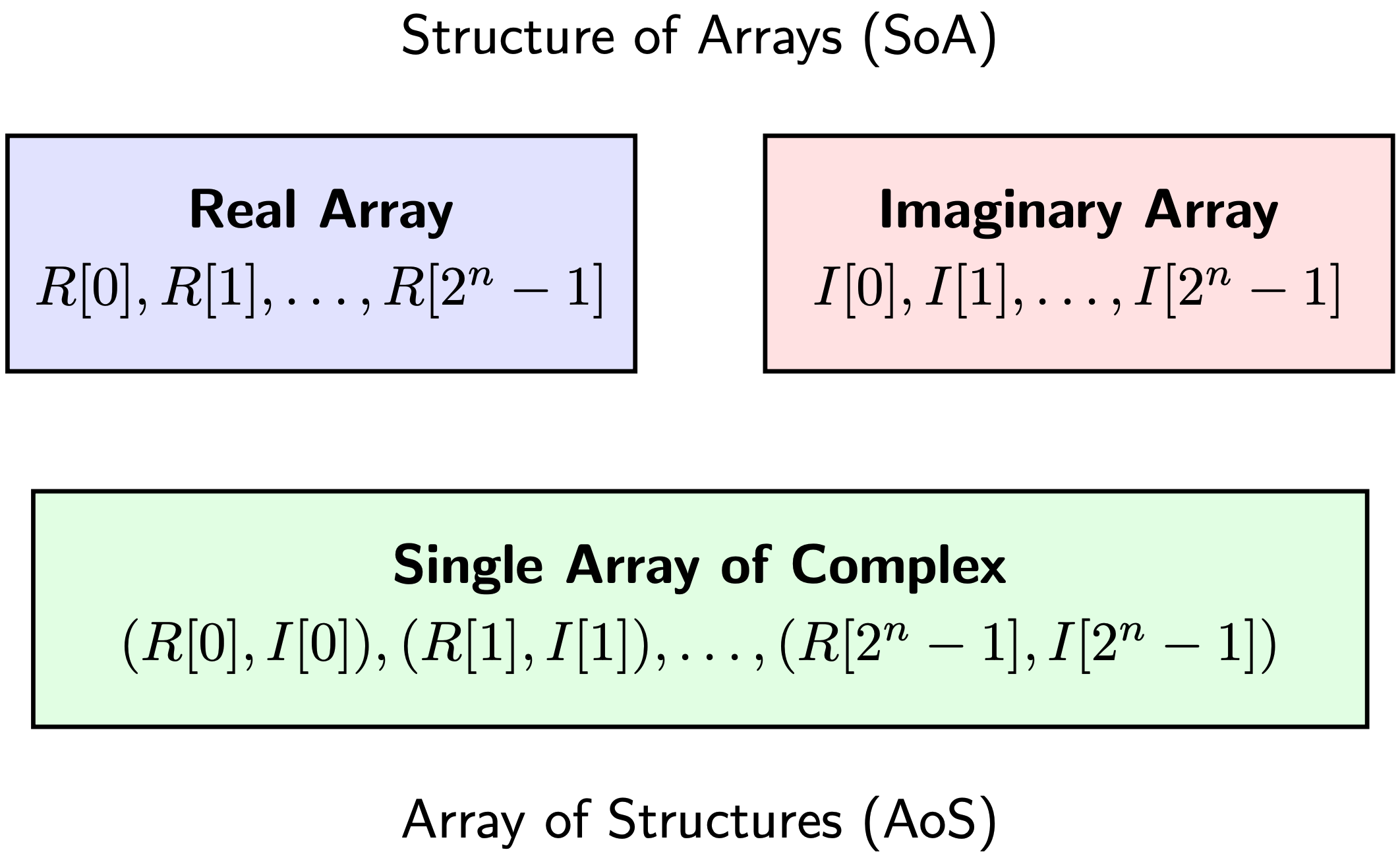}}
\caption{Schematic representation of two common data layouts for complex 
quantum amplitudes. In the Structure of Arrays (SoA) approach (top), real 
and imaginary parts are stored in separate arrays, each of length $2^n$. 
In the Array of Structures (AoS) approach (bottom), a single array of length $2^n$ 
holds both real and imaginary parts together in each element.}
\label{ds}
\end{figure}

\textbf{\textit{Today's Hardware.}}
Modern CPUs feature dozens of cores, and all state vector simulators employ multithreading to split the $2^n$ amplitude space across cores. OpenMP is commonly employed to parallelize the innermost loops of state updates. QuEST and other simulators leverage multi-core parallelism by dividing the state vector into chunks processed concurrently by threads. This yields near-linear speedup with core count~\cite{Ref4} until memory bandwidth becomes the bottleneck~\cite{Ref16}. However, multithreading on large shared-memory systems introduces challenges related to thread affinity and load balancing. If threads are not carefully pinned or if the work is unevenly distributed, performance can suffer. Thread pinning (assigning each thread to a specific core) is often required to prevent operating system (OS) scheduler overhead or migration between cores. Additionally, on multi-socket NUMA nodes, pinning helps ensure threads access memory local to their socket. While most quantum simulator papers do not explicitly describe their thread affinity policies, the need for such control is well recognized in the community~\cite{Ref17}.

On NUMA architectures, memory is physically partitioned across sockets, and accessing remote memory through the interconnect (e.g., QPI/UPI) occurs transparently, i.e., without explicit communication, but incurs higher latency and lower bandwidth. It is therefore crucial for large state vector simulations to be NUMA-aware. A straightforward strategy is to partition the state vector across NUMA nodes and pin threads such that each domain primarily works on its local portion of the data. In the ideal case, each thread only accesses the segment of the state vector residing in its local node’s memory~\cite{Ref17}. This avoids the costly scenario where a thread continuously fetches data from another socket’s memory. When thread affinity and memory placement are aligned, memory bandwidth scales with the number of NUMA nodes instead of being bottlenecked by slower cross-socket memory access~\cite{Ref18}.
Despite its importance, NUMA-aware optimization has not always been a primary focus in quantum simulator research. Many open-source simulators rely on the OS’s default first-touch policy for memory allocation (as implemented in Linux~\cite{linuxnuma}) but do not report additional NUMA-specific strategies in their publications.
We found very little to no explicit analysis of NUMA effects in simulator papers, indicating a clear gap that our work aims to address. A recent IBM patent~\cite{Ref17} describes how a naïve parallelization of a CNOT gate can lead to one socket’s thread repeatedly accessing another socket’s memory, an imbalanced situation, and proposes reordering qubit indices or replicating data to balance memory accesses. The fact that such patent-level solutions exist suggests that while NUMA issues are recognized, they have not been thoroughly explored in open literature. 

\textbf{\textit{Contributions.}}
Motivated by these observations, and building on the robust foundation of QuEST v3.7.0\footnote{Our baseline throughout this work is the default QuEST v3.7.0 build compiled with AVX2-enabled compiler optimizations (e.g., -O3 -mavx2) but without explicit intrinsics, FMA operations, or NUMA/thread placement controls. In preparing this draft, we noted that QuEST v4.0 was released, introducing several software architectural enhancements--including revised data structures that partially overlap with our improvements--but it does not incorporate explicit hand-written AVX-512 intrinsics, cache-aware loop unrolling, prefetching, and NUMA pinning/partitioning as presented in this work. Details are discussed in Section~\ref{conc}.} we present the first open-source, comprehensive evaluation of low-level and NUMA-aware optimizations for classical quantum simulation, providing a suite of single‑node enhancements forming a tunable foundation for future distributed and heterogeneous (multi-node and GPU) scaling. 
Our work consolidates and generalizes multiple architecture-specific techniques--previously studied only in isolation or without public implementations, see Table~\ref{tab:sim-optimisations}--into a single reproducible framework. 
Specifically, this paper contributes:
\begin{itemize}
    \item \textbf{(1) Open-source and reproducible low-level optimization framework:} a publicly available extension to QuEST integrating, for the first time, all key architectural optimizations--including explicit AVX-512 intrinsics, cache-aware loop unrolling, software prefetching, and NUMA-aware thread pinning--within a unified, benchmarked codebase. This enables transparent evaluation and reuse by the research community, ensuring that the improvement is not a one‑off. Many prior works were either proprietary or did not release their full code (e.g., HpQC, PAS), or they focused on a single aspect (e.g., PAS on single-node CPU, Intel-QS on distributed scaling) without covering the entire optimization spectrum.
    
    \item \textbf{(2) Fine-grained architectural tunability:} unlike most simulators that apply compiler-driven or opaque internal optimizations, our implementation exposes each optimization (e.g., AVX2 vs.\ AVX-512, loop-unrolling factor, prefetching distance, NUMA-aware vs.\ default memory allocation) as a configurable module. This tunable design enables researchers to systematically examine the impact of individual optimizations and understand their individual contribution to overall performance. Such flexibility is particularly valuable for scientific computing, as it supports reproducible and transparent performance evaluation across diverse hardware architectures. It also allows the simulator to be easily adapted to different processor generations; for example, older CPUs without AVX-512 may benefit from smaller unrolling factors or alternative memory-placement strategies, which our framework can selectively enable or disable.
    
    \item \textbf{(3) Explicit NUMA-aware execution:} we provide a detailed NUMA-partitioned memory model combined with enforced thread–memory affinity and locality-sensitive scheduling. This minimizes inter-socket memory traffic and scales memory bandwidth nearly linearly with the number of NUMA nodes--addressing a critical gap left by previous open-source simulators that assumed uniform memory access, which can leave performance on the table in multi-socket machines.
    
    \item \textbf{(4) Comprehensive performance study:} evaluating low-level optimizations across six representative workloads; single-qubit, two-qubit, Random Quantum Circuits (RQC), Quantum Fourier Transform (QFT), Grover, and Shor-like circuits, providing a holistic view of how architectural tuning affects distinct quantum workloads.
\end{itemize}

Our focus in this work is on mathematical, kernel-level architectural optimizations within QuEST rather than on full cross-simulator benchmarking. We systematically examine how different low-level optimizations affect performance within a single software framework. Prior studies, including the original QuEST paper~\cite{Ref4} and recent Queen quantum simulator \cite{queen2024}, have consistently shown QuEST to be competitive or superior to simulators such as Qiskit Aer, Intel-QS, cuQuantum, and qsim across a wide range of architectures and workloads. This supports our choice of QuEST as a representative high-performance baseline. 
We specifically target node-local memory bottlenecks--an often overlooked yet critical aspect of quantum simulation performance. Similar to PAS, HpQC, and Qulacs, we begin with single-node optimization as a necessary foundation for scalable multi-node execution. We prioritize deep, low-level kernel tuning over early MPI distribution, since premature scaling without resolving intra-node memory and threading bottlenecks can degrade overall performance. By first addressing locality and vectorization challenges, we establish a robust baseline for subsequent distributed extensions, which are under active development and will be presented in a follow-up study.

\textbf{\textit{Paper Organization.}}
Section~\ref{methodology} details our system architecture and optimization methodology—including our novel data structure layout (Section~\ref{DSEnhansement}), NUMA-aware optimization strategies (Sections~\ref{sec3B} and~\ref{TA}), and various instruction-level tuning techniques (Section~\ref{secTuning}). Section~\ref{SecD} presents our benchmark results and analysis, covering the performance of single- and two-qubit gate operations, QFT, RQC, Grover, and a Shor-like periodicity estimation circuit. As demonstrated, integrating our newly designed data layout and these targeted optimizations yields performance gains comparable to those reported by HpQC and PAS, although not as pronounced as some of their most optimistic findings. Section~\ref{disc} discusses the scaling and portability of our framework. Finally, Section~\ref{conc} concludes with a summary of our findings and a discussion of potential directions for future research. 
\begin{figure}[!htbp]
\centerline{\includegraphics[width=0.5\textwidth]{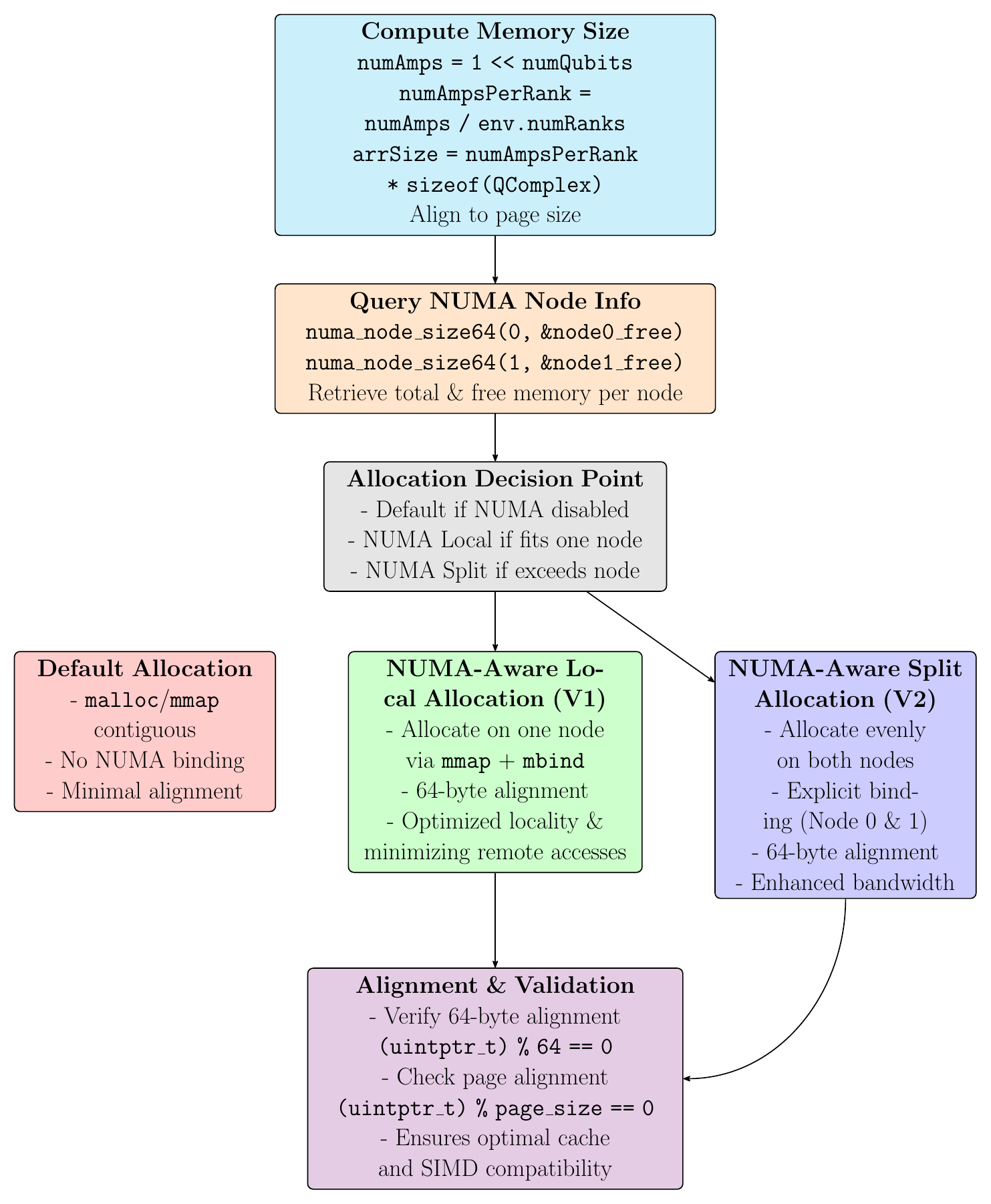}}
\caption{Comparison of three state vector allocation strategies on our dual-node NUMA system.}
\label{alloc}
\end{figure}
\begin{figure*}[!htbp]
\centerline{\includegraphics[width=\linewidth]{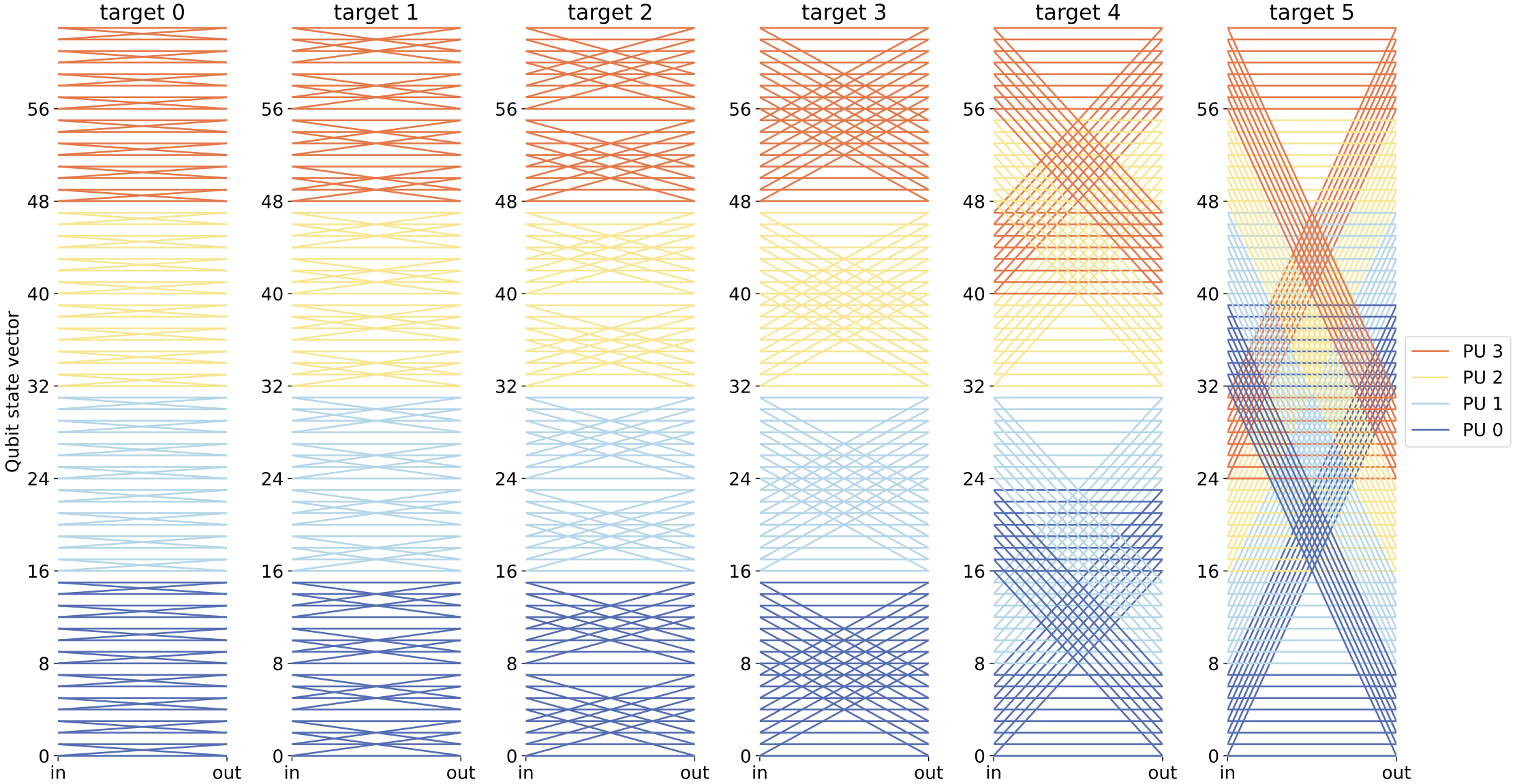}}
\caption{Access pattern for in‐place matrix‐vector multiplication with single‐qubit (unitary) gates on 6 qubits. Each sub‐plot (target 0 to 5) shows how amplitude indices (vertical axis) are reordered in memory during gate application. Colored lines map amplitude blocks to four parallel CPU cores (PU 0–3).}
\label{unitary6q}
\end{figure*}
\begin{figure*}[htbp]
\centerline{\includegraphics[width=0.9\linewidth]{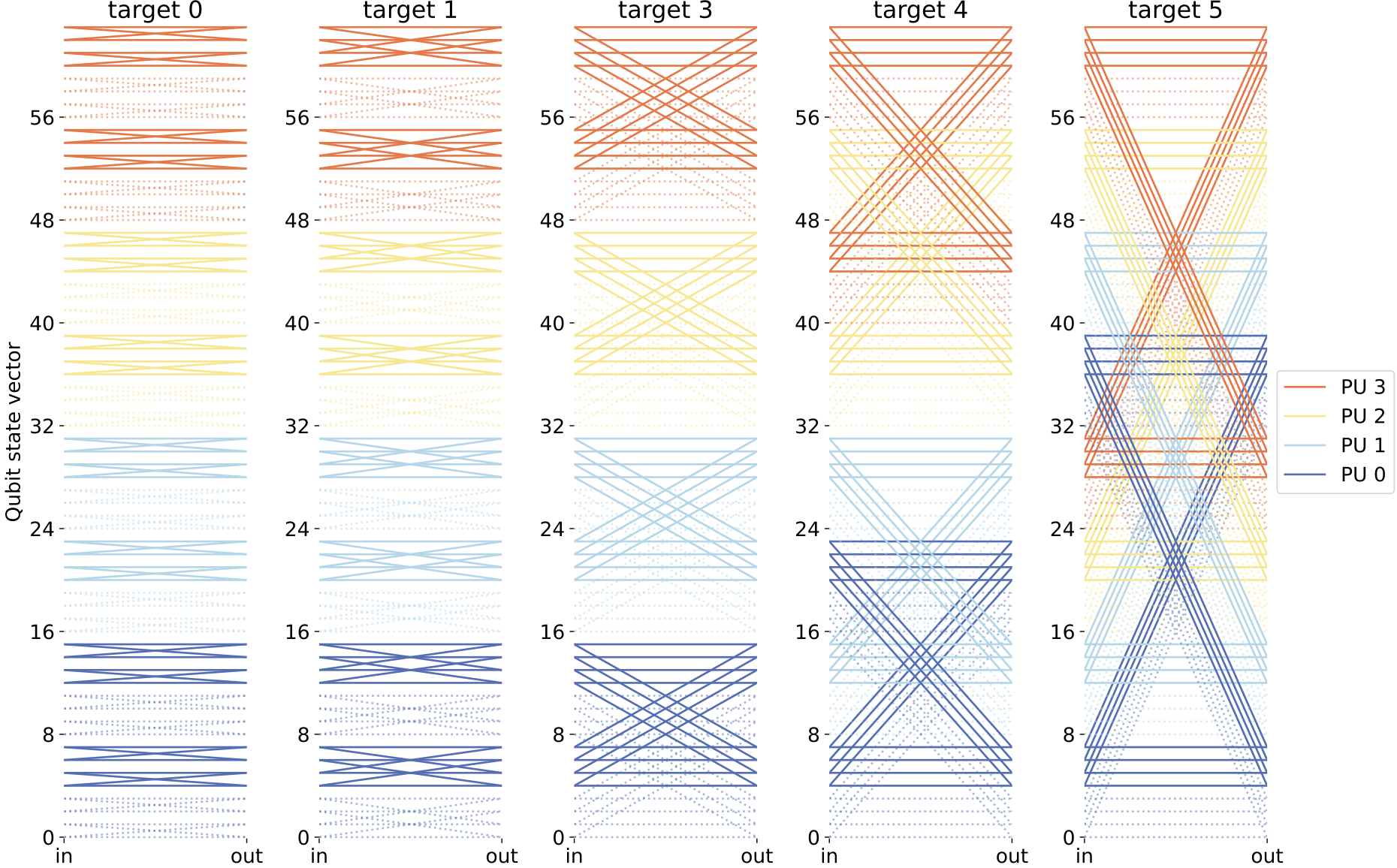}}
\caption{Access pattern for in-place matrix-vector multiply with controlled gates on 6 qubits, using qubit 2 as the control. Solid lines represent amplitudes updated (control bit = 1), while dashed lines indicate skipped computations (control bit = 0).}
\label{control6q2}
\end{figure*}
\section{System Architecture and Methodology} \label{methodology}

\subsection{Data Structure Enhancements} \label{DSEnhansement}
In the baseline QuEST v3.7.0 implementation, complex amplitudes are stored in a structure-of-arrays (SoA) format: a \texttt{ComplexArray} with two separate arrays, one for all real parts and one for all imaginary parts. While this representation simplifies certain operations, it leads to inefficient memory access and poor cache utilization because accessing a single complex amplitude requires touching two distant memory locations. We refactored this into an array-of-structures (AoS) layout, packing each amplitude’s real and imaginary components contiguously in memory. This AoS format (illustrated in Figure~\ref{ds}) places the real and imaginary parts of each amplitude next to each other, improving spatial locality. As a result, when iterating through the state vector, each cache line fetched contains complete complex values, reducing cache misses and better exploiting hardware prefetching. We also ensure that the allocated array of complex numbers is aligned in memory (64-byte alignment), which facilitates aligned vector load/store operations in later optimizations. This transition from an AoS to a SoA improves effective cache use and SIMD efficiency in large-scale quantum circuit simulations, where large arrays of complex numbers are processed frequently.

\subsection{NUMA-Aware Optimization}
Modern multi-socket servers employ NUMA, where each socket has local memory with higher bandwidth and lower latency than remote memory connected through inter-socket links such as UPI or Infinity Fabric. 
Without NUMA-awareness--i.e., if left to the default OS policies--the memory of a process may be spread across sockets in a suboptimal way, and threads may migrate between nodes, leading to costly remote memory accesses. 
In multi-threaded simulations such as QuEST, these effects can severely limit scaling efficiency if threads frequently access data residing on another socket.
While in distributed-memory (MPI-based) settings this issue is often mitigated by binding MPI ranks to NUMA regions, our focus here is on single-process, multi-threaded execution where such process-level pinning is not directly applicable. 
Therefore, in this section, we employ two complementary strategies to optimize memory placement and thread affinity on a dual- (multi-) NUMA system.

\subsubsection{Memory Allocation} \label{sec3B}
By default, QuEST allocates the state vector with the standard libc malloc, which on Linux follows a first-touch policy for NUMA (pages are allocated on the node where they are first accessed).
This means that the OS is free to distribute pages across NUMA nodes arbitrarily, which may not be optimal on systems with multiple memory nodes, leading to performance penalties from remote memory accesses.
To address this limitation, we introduce explicit control over memory placement via two NUMA-optimized strategies: (i) Local Allocation (V1), where the code first queries the free memory on each NUMA node (via \texttt{numa\_node\_size64}), then checks if the required memory can fit into a single node. If so, it performs an \texttt{mmap} allocation and sets a bind policy with \texttt{mbind} to restrict that allocation strictly to the chosen node. This ensures localized memory access, maximizing the benefit of low-latency local memory. If the memory size exceeds what one node can provide, it gracefully falls back to partial allocation on the other node, while still enforcing page alignment. (ii) Split Allocation (V2), where we always split the allocated memory evenly across two nodes from the start. After computing halfSize (and ensuring that it remains at least one page in size), we allocate the entire region via \texttt{mmap}, then explicitly bind the first half of the region to Node 0 and the second half to Node 1. This approach ensures a balanced distribution of memory which yields higher total memory bandwidth. As in V1, the routine first rounds the requested buffer size up to an exact multiple of the system page size, allocates it with \texttt{mmap}, and verifies that the returned pointer is 64-byte aligned--guaranteeing efficient AVX-512 loads and stores. Both strategies confirm that the base address still starts on a page boundary, eliminating sub-page fragmentation that could otherwise hamper memory-access performance. Figure~\ref{alloc} compares the available allocation strategies schematically. These methods integrate seamlessly within the QuEST environment, allowing the user to switch between Default, NUMA V1, or NUMA V2 at compile or runtime, depending on the specific memory requirements and system architecture. 
 \begin{algorithm}[H]
    \caption{Hadamard kernel (Default QuEST v3.7.0 with the new AoS data layout)}
    \label{alg:hadamardLocalDefault}
    \begin{algorithmic}[1]
        \REQUIRE \texttt{qureg} (with \texttt{stateVec.elements}), integer \texttt{targetQubit}
        \ENSURE Updates \texttt{qureg.stateVec.elements} in-place with the Hadamard transform on \texttt{targetQubit}

        \STATE $sizeHalfBlock \gets 1 \ll \texttt{targetQubit}$
        \STATE $sizeBlock \gets 2 \times sizeHalfBlock$
        \STATE $numTasks \gets \texttt{qureg.numAmpsPerChunk} \gg 1$
        \STATE $\texttt{recRoot2} \gets 1/\sqrt{2}$
        \STATE Let \texttt{stateVecElements} be a pointer to \texttt{qureg.stateVec.elements}

        \STATE \textbf{OpenMP parallel region}:
        \FOR{$\texttt{thisTask} \gets 0$ \textbf{to} $numTasks - 1$}
            \STATE $\texttt{thisBlock} \gets \lfloor \texttt{thisTask} / \texttt{sizeHalfBlock} \rfloor$
            \STATE $\texttt{indexUp} \gets \texttt{thisBlock} \times \texttt{sizeBlock} + (\texttt{thisTask} \bmod \texttt{sizeHalfBlock})$
            \STATE $\texttt{indexLo} \gets \texttt{indexUp} + \texttt{sizeHalfBlock}$

            \STATE $\texttt{stateRealUp} \gets$\\
            $\quad \texttt{stateVecElements}[\texttt{indexUp}].\texttt{real}$

            \STATE $\texttt{stateImagUp} \gets$\\
            $\quad \texttt{stateVecElements}[\texttt{indexUp}].\texttt{imag}$

            \STATE $\texttt{stateRealLo} \gets$\\
            $\quad \texttt{stateVecElements}[\texttt{indexLo}].\texttt{real}$

            \STATE $\texttt{stateImagLo} \gets$\\
            $\quad \texttt{stateVecElements}[\texttt{indexLo}].\texttt{imag}$

            \STATE $\texttt{stateVecElements}[\texttt{indexUp}].\texttt{real} \gets \texttt{recRoot2} \times (\texttt{stateRealUp} + \texttt{stateRealLo})$
            \STATE $\texttt{stateVecElements}[\texttt{indexUp}].\texttt{imag} \gets \texttt{recRoot2} \times (\texttt{stateImagUp} + \texttt{stateImagLo})$
            \STATE $\texttt{stateVecElements}[\texttt{indexLo}].\texttt{real} \gets \texttt{recRoot2} \times (\texttt{stateRealUp} - \texttt{stateRealLo})$
            \STATE $\texttt{stateVecElements}[\texttt{indexLo}].\texttt{imag} \gets \texttt{recRoot2} \times (\texttt{stateImagUp} - \texttt{stateImagLo})$
        \ENDFOR
    \end{algorithmic}
\end{algorithm}

\subsubsection{Thread Affinity} \label{TA}
Locality‐aware task scheduling is essential in NUMA architectures, as it ensures that threads execute on the memory node where their data reside, thereby minimizing costly remote accesses. In our approach, we dynamically calculate task chunks and bind threads explicitly to NUMA nodes rather than relying on the OS's scheduler.
Half of the threads are pinned to the first socket, and the other half to the second socket--aligned with the layout used in allocation V2 (see Algorithm~\ref{alg:hadamard_full_ref}). 
The state vector workload is divided accordingly so that each thread operates primarily on the portion of the state vector stored in its local node’s memory. 
In our two-socket system, threads $0…N/2-1$ process the lower half of the amplitude array (allocated on node~0), while threads $N/2…N-1$ process the upper half (allocated on node~1). Algorithm~\ref{alg:numa_affinity_simple} illustrates the NUMA-aware thread binding strategy used in our implementation. 
During initialization, each OpenMP thread determines its global ID and binds itself to the corresponding NUMA node, ensuring memory locality by keeping both the working set and execution context within the same NUMA domain and thereby reducing cross-node memory latency.

\begin{algorithm}[H]
\caption{NUMA-aware thread affinity}
\label{alg:numa_affinity_simple}
\begin{algorithmic}[1]
\REQUIRE $N_{\text{threads}},\; N_{\text{nodes}} = 2$
\ENSURE Each thread is bound to the closest NUMA node
\FOR{each OpenMP thread $t_i$}
    \STATE $tid \gets$ \Call{GetThreadID}{}
    \STATE $node \gets 
        \begin{cases}
            0, & \text{if } tid < N_{\text{threads}}/2 \\
            1, & \text{otherwise}
        \end{cases}$
    \STATE \Call{BindThreadToNode}{$t_i, node$}
\ENDFOR
\end{algorithmic}
\end{algorithm}
This manual partitioning and pinning strategy maximizes data locality for most workloads and prevents idle resources when tasks are not perfectly divisible. Combined with optimizations such as loop unrolling and cache‐friendly data layouts, it significantly lowers memory‐access latency and boosts overall performance.

\begin{figure}[!htbp]
\centerline{\includegraphics[width=0.5\textwidth]{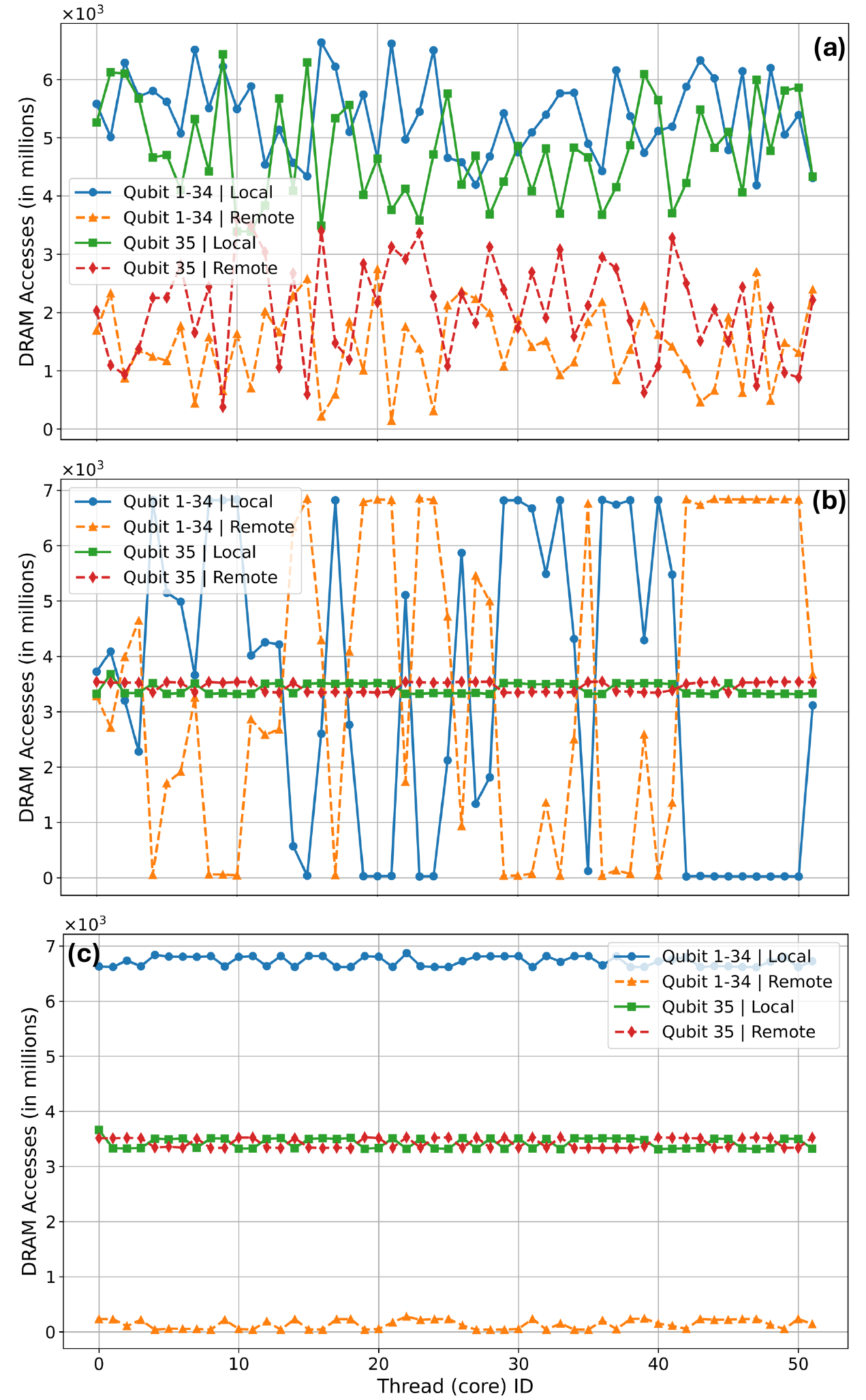}}
\caption{Comparison of DRAM access for 50 consecutive Hadamard gates measured with Intel PCM on a dual-NUMA system. (a) Baseline QuEST implementation (default v3.7.0) shows unbalanced memory traffic across sockets; (b) NUMA-aware memory allocation reduces remote accesses by improving locality; and (c) further optimization using a locality-sensitive task scheduler yields near-ideal confinement of DRAM activity to local nodes. Here, the full state vector fits within a single NUMA node up to 34 qubits, while 35 qubits and above require cross-node memory access.}
\label{ph4}
\end{figure}

Figure~\ref{ph4} evaluates the impact of NUMA-aware optimizations on a sequence of 50 Hadamard gates, profiled with Intel PCM NUMA~\cite{Ref19}. 
The Y-axis shows DRAM accesses observed by individual threads (on the X-axis) for different target qubits. Configuration (a) represents the baseline QuEST implementation, exhibiting scattered and suboptimal memory accesses across nodes. 
Configuration (b) introduces NUMA-aware memory allocation while retaining the baseline Hadamard kernel, thereby improving locality by favoring accesses to local memory. 
Configuration (c) further refines this approach by employing a locality‐sensitive scheduler. As a result, when the working set fits entirely within a single NUMA node--up to 34 qubits on our test system--all DRAM accesses remain local, effectively minimizing remote memory traffic. When the state vector spans multiple NUMA nodes (e.g., at 35 qubits), a mix of local and remote accesses occurs, yet cross-node traffic remains significantly reduced. 
Comparable improvements can also be achieved using environment variables (e.g., \texttt{OMP\_PROC\_BIND}, \texttt{OMP\_PLACES}, \texttt{GOMP\_CPU\_AFFINITY}) to enforce core affinity, confirming the critical role of thread binding in reducing latency.
However, manually toggling and tuning these environment variables can be error-prone and tedious, especially across systems with different NUMA topologies. 
In contrast, our implementation performs this binding automatically based on the detected NUMA topology, while still allowing users to override or adjust the mapping policy at compile time or runtime if needed.

\subsection{Instruction-level Tuning} \label{secTuning}
With the optimized data layout and NUMA strategy established, we further enhance the innermost simulation loops through instruction-level optimizations that exploit fine-grained hardware parallelism.
\begin{algorithm}[H]
\caption{Vectorized Hadamard using AVX-512 intrinsics} 
\label{alg:avx512_hadamard}
\begin{algorithmic}[1]
\REQUIRE State vector $\psi$, target qubit index $q_t$
\ENSURE Apply Hadamard transform on amplitudes paired across $q_t$
\STATE $r_2 \gets 1/\sqrt{2}$ \COMMENT{Normalization constant}
\FOR{each block of amplitude pairs $(\psi_u, \psi_l)$}
    \STATE \textbf{Prefetch} $\psi_u, \psi_l$ into cache
    \STATE $\mathbf{up} \gets$ \Call{Load512}{$\psi_u$}
    \STATE $\mathbf{lo} \gets$ \Call{Load512}{$\psi_l$}
    \STATE $\mathbf{lo_s} \gets r_2 \times \mathbf{lo}$
    \STATE $\mathbf{newUp} \gets r_2 \times \mathbf{up} + \mathbf{lo_s}$
    \STATE $\mathbf{newLo} \gets r_2 \times \mathbf{up} - \mathbf{lo_s}$
    \STATE \Call{Store512}{$\psi_u$, $\mathbf{newUp}$}
    \STATE \Call{Store512}{$\psi_l$, $\mathbf{newLo}$}
\ENDFOR
\end{algorithmic}
\end{algorithm}
\subsubsection{Vectorization with AVX-512 Intrinsics}
Vectorization enables a processor to exploit data-level parallelism by performing identical operations on multiple data elements within a single instruction cycle.
This often yields substantial performance gains for applications that repeatedly apply the same operation across large datasets.
In state vector simulation, the core operation--a matrix-vector multiplication--exhibits strong data parallelism due to the regular and sparse structure of quantum gate matrices.

However, relying solely on compiler auto-vectorization is insufficient to achieve meaningful speedups because of the way the original QuEST code is written.
In particular, the gate-application loops in QuEST~v3.7.0 rely heavily on bitwise operations, in particular masking and shifting, to identify which iterations to skip for controlled gates and to compute the corresponding memory indices from the iteration index.
Such irregular access patterns do not translate efficiently into vectorized instructions. 
Moreover, even when vectorization succeeds and instructions are issued at higher throughput, overall performance remains limited by memory bandwidth, which prevents these gains from directly translating into faster gate execution times.

\begin{algorithm}[H]
\caption{AVX-512 Hadamard mini-kernel (4 complex pairs)}
\label{alg:avx512_microkernel}
\begin{algorithmic}[1]
\REQUIRE Registers $\mathbf{up}, \mathbf{lo}$ loaded from memory, broadcast $\mathbf{r_2}$
\ENSURE Updated $\mathbf{up}, \mathbf{lo}$ stored back in place
\STATE $\mathbf{lo_s} \gets \mathbf{r_2} \times \mathbf{lo}$  \COMMENT{scale lower amplitudes}
\STATE $\mathbf{newUp} \gets \Call{FMA}{\mathbf{r_2}, \mathbf{up}, \mathbf{lo_s}}$ \COMMENT{$\mathbf{r_2} \cdot \mathbf{up} + \mathbf{lo_s}$}
\STATE $\mathbf{newLo} \gets \Call{FMS}{\mathbf{r_2}, \mathbf{up}, \mathbf{lo_s}}$ \COMMENT{$\mathbf{r_2} \cdot \mathbf{up} - \mathbf{lo_s}$}
\STATE \Call{Store512}{$\mathbf{newUp}$, $\mathbf{newLo}$} \COMMENT{write back to $\psi$}
\end{algorithmic}
\end{algorithm}

In our implementation, we explicitly write the optimized kernels using intrinsic AVX-512 instructions, which perform simultaneous operations on eight double-precision floating-point values (corresponding to four complex amplitudes). 
By using aligned vector loads and stores (via \texttt{\_mm512\_load\_pd} and \texttt{\_mm512\_store\_pd}), the code minimizes memory-transfer overhead. 
Arithmetic operations are further optimized with FMA instructions, which combine multiplication and addition (or subtraction) in a single operation--common in quantum gate updates--thereby further reducing the instruction count and improving numerical accuracy.

Algorithm~\ref{alg:avx512_hadamard} outlines the high-level AVX-512 vectorization strategy used for the Hadamard kernel. Each loop iteration processes four amplitude pairs simultaneously (eight complex double-precision numbers). The intrinsic operations \texttt{\_mm512\_load\_pd}, \texttt{\_mm512\_fmadd\_pd}, and \texttt{\_mm512\_fmsub\_pd} are abstracted here as high-level vector load, fused multiply-add, and store operations, enabling efficient SIMD parallelism with minimal memory latency.

While Algorithm~\ref{alg:avx512_hadamard} presents the vectorization at a high level,
the actual compute loop is implemented as a dedicated AVX-512 mini-kernel that fuses multiplication and addition/subtraction within a single instruction for maximum throughput.
This mini-kernel is repeatedly invoked from the main optimized kernel (Algorithm~\ref{alg:hadamard_full_ref}).
\begin{algorithm}[H]
\caption{Loop unrolling with pointer restriction}
\label{alg:loop_unrolling}
\begin{algorithmic}[1]
\REQUIRE State vector $\psi$, unroll factor $U = 16$
\ENSURE Reduced loop overhead and increased ILP
\STATE \texttt{stateVec} $\gets$ \texttt{restrict pointer to } $\psi$
\FOR{$i = 0$ \textbf{to} $N$ \textbf{step} $U$}
    \FOR{$u = 0$ \textbf{to} $U-1$}
        \STATE process amplitude pair $(\psi_{i+u}, \psi_{i+u+sizeBlock})$
    \ENDFOR
\ENDFOR
\end{algorithmic}
\end{algorithm}

\subsubsection{Loop Unrolling and Pointer Restriction}
Aggressive loop unrolling is employed--processing blocks of 16 iterations per step--to reduce the overhead of loop control and enhance instruction-level parallelism on the CPU. The use of the \texttt{restrict} qualifier further assists the compiler by confirming non-aliasing pointers, enabling more aggressive vectorization and other low-level optimizations, as illustrated in Algorithm~\ref{alg:loop_unrolling}.
\begin{algorithm}[H]
\caption{NUMA-aware workload partitioning}
\label{alg:numa_partition}
\begin{algorithmic}[1]
\REQUIRE State vector $\psi$ of size $S$, $N_{\text{nodes}} = 2$, $N_{\text{threads}}$
\ENSURE Partition the workload to maximize memory locality
\STATE $chunkSize \gets S / N_{\text{nodes}}$
\FOR{$n = 0$ \textbf{to} $N_{\text{nodes}} - 1$}
    \STATE $\textit{start}[n] \gets n \times chunkSize$
    \STATE $\textit{end}[n] \gets (n+1) \times chunkSize - 1$
    \STATE \Call{AllocateMemoryOnNode}{$\psi[\textit{start}[n] : \textit{end}[n]]$, $n$}
\ENDFOR
\FOR{each thread $t_i$}
    \STATE $tid \gets$ \Call{GetThreadID}{}
    \STATE $node \gets$ \Call{NodeOfThread}{$tid$}
    \STATE \Call{ProcessBlock}{$\psi[\textit{start}[node] : \textit{end}[node]]$}
\ENDFOR
\end{algorithmic}
\end{algorithm}

\subsubsection{Memory Optimization and NUMA Awareness}
In addition to vector-level improvements, the implementation incorporates memory optimization strategies that are critical for high-performance execution. Utilising OpenMP, each thread operates on a distinct segment of the state vector, and a NUMA-aware approach is adopted by binding threads to specific NUMA nodes (using \texttt{numa\_run\_on\_node}) to mitigate cross-node memory latencies and optimize memory locality. 

In addition to thread binding, we explicitly partition the state vector between NUMA nodes to ensure memory is allocated close to the threads that will access it.
As illustrated in Algorithm \ref{alg:numa_partition}, the state vector is divided into contiguous chunks, with each chunk allocated on the memory node corresponding to a specific NUMA domain. Threads pinned to a node process only their local chunk of the state vector, ensuring that nearly all memory accesses remain local.
This NUMA-aware partitioning works synergistically with thread affinity (Algorithm \ref{alg:numa_affinity_simple}): the former places data where it belongs, while the latter ensures computation happens near the data. Together, these reduce remote DRAM accesses and improve effective bandwidth utilization.
When the state vector fits within a single NUMA domain, all accesses are local; as the problem scales across nodes, this partitioning strategy minimizes—but cannot entirely eliminate—remote memory traffic.
\begin{algorithm}[H]
\caption{Memory prefetching for vectorized loop}
\label{alg:prefetch}
\begin{algorithmic}[1]
\REQUIRE Prefetch distances $d_1 = 16$, $d_2 = 32$
\ENSURE Lower memory access latency during vector ops
\FOR{each vectorized iteration $i$}
    \STATE \Call{Prefetch}{$\psi[i + d_1]$}
    \STATE \Call{Prefetch}{$\psi[i + d_2]$}
    \STATE \Call{Prefetch}{$\psi[i + sizeBlock + d_1]$}
    \STATE \Call{Prefetch}{$\psi[i + sizeBlock + d_2]$}
    \STATE process 4 amplitude pairs using AVX-512
\ENDFOR
\end{algorithmic}
\end{algorithm}

\subsubsection{Memory Prefetching}
To further reduce memory access latency, for each group of 4 complex numbers processed, the code explicitly employs prefetching via \texttt{\_mm\_prefetch} to load future cache lines in advance, reducing cache miss penalties when processing large blocks of data. This is critical for wide vector loads (\texttt{\_mm512\_load\_pd}) to sustain bandwidth. Our kernel processes eight doubles at a time, hence prefetching data 16 and 32 elements ahead ensures that the subsequent iterations’ data is already loaded into the L1 cache.
\begin{algorithm}[H]
\caption{Scalar fallback for remaining amplitudes}
\label{alg:scalar_fallback}
\begin{algorithmic}[1]
\REQUIRE Remaining elements $R < \text{vector width}$
\ENSURE Full correctness of transformation
\FOR{each remaining amplitude $r \in R$}
    \STATE \Call{Prefetch}{$\psi[r+1]$}
    \STATE \Call{Prefetch}{$\psi[r+sizeBlock+1]$}
    \STATE compute scalar Hadamard:
    \STATE $\psi'_u = r_2(\psi_u + \psi_l)$,\quad $\psi'_l = r_2(\psi_u - \psi_l)$
\ENDFOR
\end{algorithmic}
\end{algorithm}

\subsubsection{Scalar Fallback for Complete Coverage}
To guarantee complete processing of the state vector, any iterations that cannot be handled by the vectorized loop are managed via a scalar fallback routine. This ensures that all amplitudes are accurately transformed, even when they do not align with the vectorized block sizes. A similar prefetching (with a smaller offset) ensures even the remaining, non-vectorized elements benefit from lower cache miss penalties.

These optimizations are illustrated in pseudocode: Algorithm~\ref{alg:hadamardLocalDefault} shows the baseline Hadamard transformation incorporating the new AoS, while Algorithm~\ref{alg:hadamard_full_ref} shows our fully optimised kernel. This implementation is engineered to maximize throughput on modern multi-core architectures through a series of low-level optimizations\footnote{Our AVX-512 implementation goes beyond simply widening AVX2 with FMA instructions. We employ a 512-bit aligned SoA layout, unroll loops in blocks of 16 iterations (corresponding to 4 complex amplitudes), and apply fixed-distance prefetching (e.g., +16, +32 elements) to reduce L1 cache latency. These elements, combined with carefully hand-written FMA-based vector kernels, significantly improve cache utilization and instruction throughput compared to the AVX2-AoS baseline.
} presented in Section~\ref{methodology}. 
\begin{algorithm}[H]
    \caption{Fully Optimized Hadamard Kernel}
    \label{alg:hadamard_full_ref}
    \begin{algorithmic}
        \REQUIRE Target qubit $q_t$, state vector $\psi$ of length $N$
        \ENSURE Apply Hadamard transform with all optimizations

        \STATE $sizeBlock \gets 1 \ll q_t$
        \STATE $numBlocks \gets \lceil N / (2 \times sizeBlock) \rceil$
        \STATE $stateVec \gets$ \texttt{restrict pointer to } $\psi$
        \STATE $r_2 \gets 1/\sqrt{2}$; \hspace{1em} $\mathbf{r_2} \gets$ broadcast of $r_2$ to AVX-512 register

        \STATE \textbf{OpenMP parallel region:}
        \STATE \quad \Call{NUMAThreadBind}{threads, nodes} \COMMENT{Algorithm~\ref{alg:numa_affinity_simple}}
        \STATE \quad \Call{NUMAPartition}{stateVec, nodes} \COMMENT{Algorithm~\ref{alg:numa_partition}}

        \FOR{$block = 0$ \textbf{to} $numBlocks - 1$}
            \STATE $indexStart \gets block \times 2 \times sizeBlock$
            \STATE $i \gets 0$
            \STATE \textbf{while} $i \le sizeBlock - U$ \textbf{do} \COMMENT{Loop unrolling (Alg.~\ref{alg:loop_unrolling})}
            \STATE \quad \textbf{for} $j = 0$ \textbf{to} $U - 1$ \textbf{step}
            $4$ \textbf{do}
            \STATE \quad\quad \Call{Prefetch}{$\psi[i + j + d]$} \COMMENT{Alg.~\ref{alg:prefetch}}
            \STATE \quad\quad \Call{AVX512\_FMA\_MiniKernel}{$\psi[i + j], \psi[i + j + sizeBlock]$} \COMMENT{Alg.~\ref{alg:avx512_microkernel}}
            \STATE \quad \textbf{end for}; \quad $i \gets i + U$
            \STATE \textbf{end while}
            \STATE \textbf{while} $i \le sizeBlock - 4$ \textbf{do}
            \STATE \quad \Call{AVX512\_FMA\_MiniKernel}{$\psi[i], \psi[i + sizeBlock]$}; \hspace{0.3em} $i \gets i + 4$
            \STATE \textbf{end while}
            \STATE \textbf{while} $i < sizeBlock$ \textbf{do}
            \STATE \quad \Call{ScalarFallback}{$\psi[i], \psi[i + sizeBlock]$} \COMMENT{Alg.~\ref{alg:scalar_fallback}}
            \STATE \quad $i \gets i + 1$
            \STATE \textbf{end while}
        \ENDFOR
    \end{algorithmic}
\end{algorithm}


\section{Benchmark Results and Analysis} \label{SecD}
The experiments were conducted on a dual-socket Intel® Xeon® Gold 6230R system (x86\_64) featuring 52 physical cores (26 per socket), with hyperthreading disabled to ensure consistent core performance. Each core includes 32 KiB of instruction and 32 KiB of data L1 cache, and 1 MiB of L2 cache. Each socket includes 35.75MiB L3 cache, for a total of 71.5 MiB amongst the two sockets.
The system is configured with two NUMA nodes, each offering approximately 385 GB of local memory. Notably, local memory latency is about 100 ns, while remote memory latency reaches 150 ns, as measured using Intel's MLC, highlighting the importance of NUMA-aware optimizations.
To ensure stable and reproducible performance results, we disabled Intel Turbo Boost and turned off CPU frequency scaling, preventing transient clock speed variations from influencing the benchmarks. All experiments were performed on a Linux environment, and the code was compiled in Release mode with GCC 9.3.0 (C99, C++98) via CMake 3.7, employing aggressive optimization flags (e.g., -O3) to fully leverage the hardware’s performance capabilities. Moreover, for consistent benchmarking, we utilized Google Benchmark integrated within the Yao framework~\cite{Ref20}, where to enhance statistical reliability each benchmark operation (i.e., an individual gate or a composite circuit) was repeated ten times and the minimum simulation time is reported.
\begin{figure}[htbp]
\centerline{\includegraphics[width=0.5\textwidth]{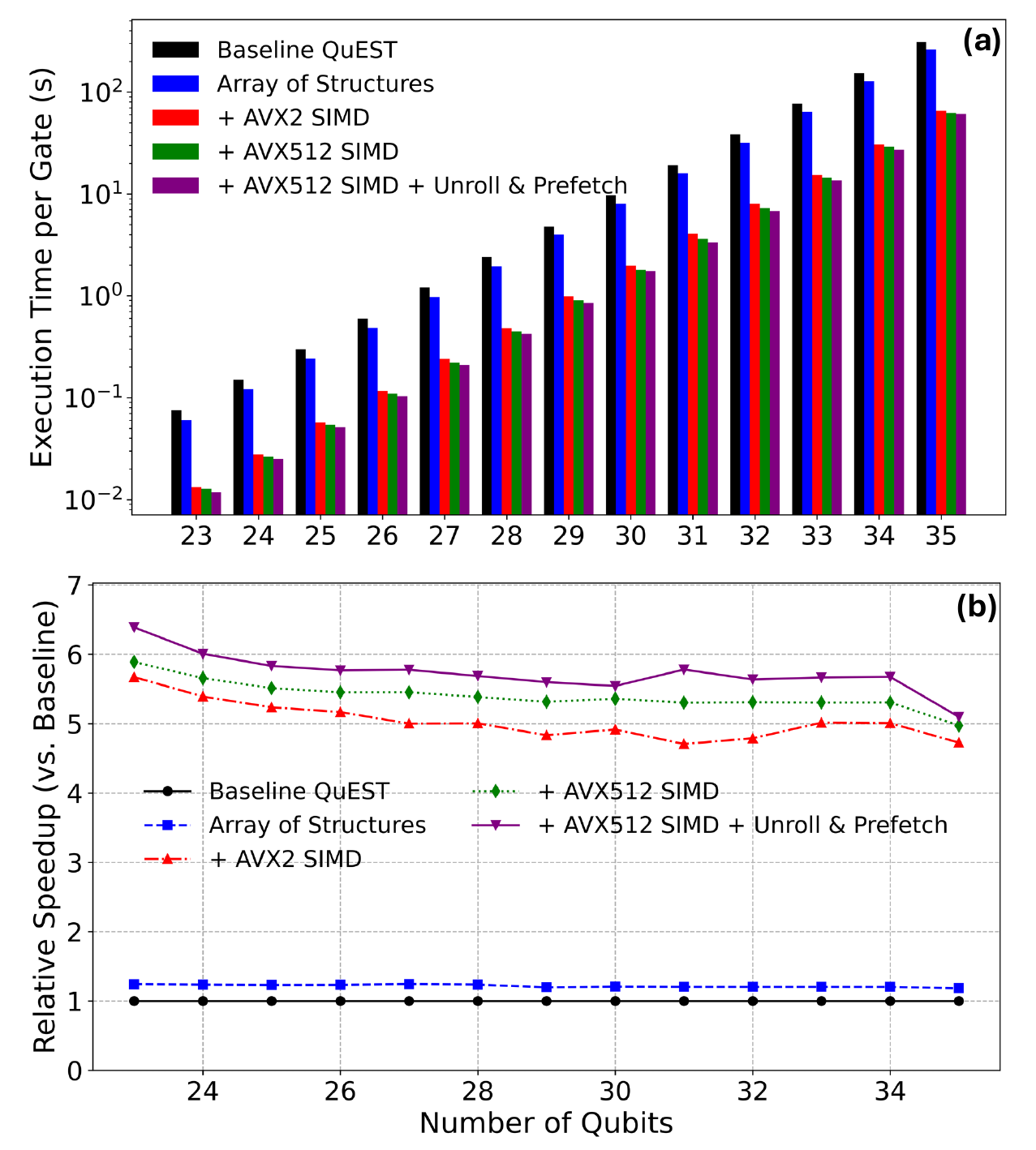}}
\caption{(a) Per-gate execution time (s) for Hadamard gate applied on the second qubit of register ranging from 23 up to 35 qubits. Bars show the baseline QuEST (default v3.7.0 build, in black), new AoS data layout (blue), AoS + intrinsic AVX2 SIMD (red), AoS + AVX-512 SIMD (green), and AVX-512 SIMD with AoS, FMA, loop unrolling \& prefetching (purple).
(b) Relative Speedup for the same set of optimizations, highlighting how each technique improves performance over the baseline QuEST implementation.}
\label{ben1}
\end{figure}

\begin{figure}[htbp]
\centerline{\includegraphics[width=0.5\textwidth]{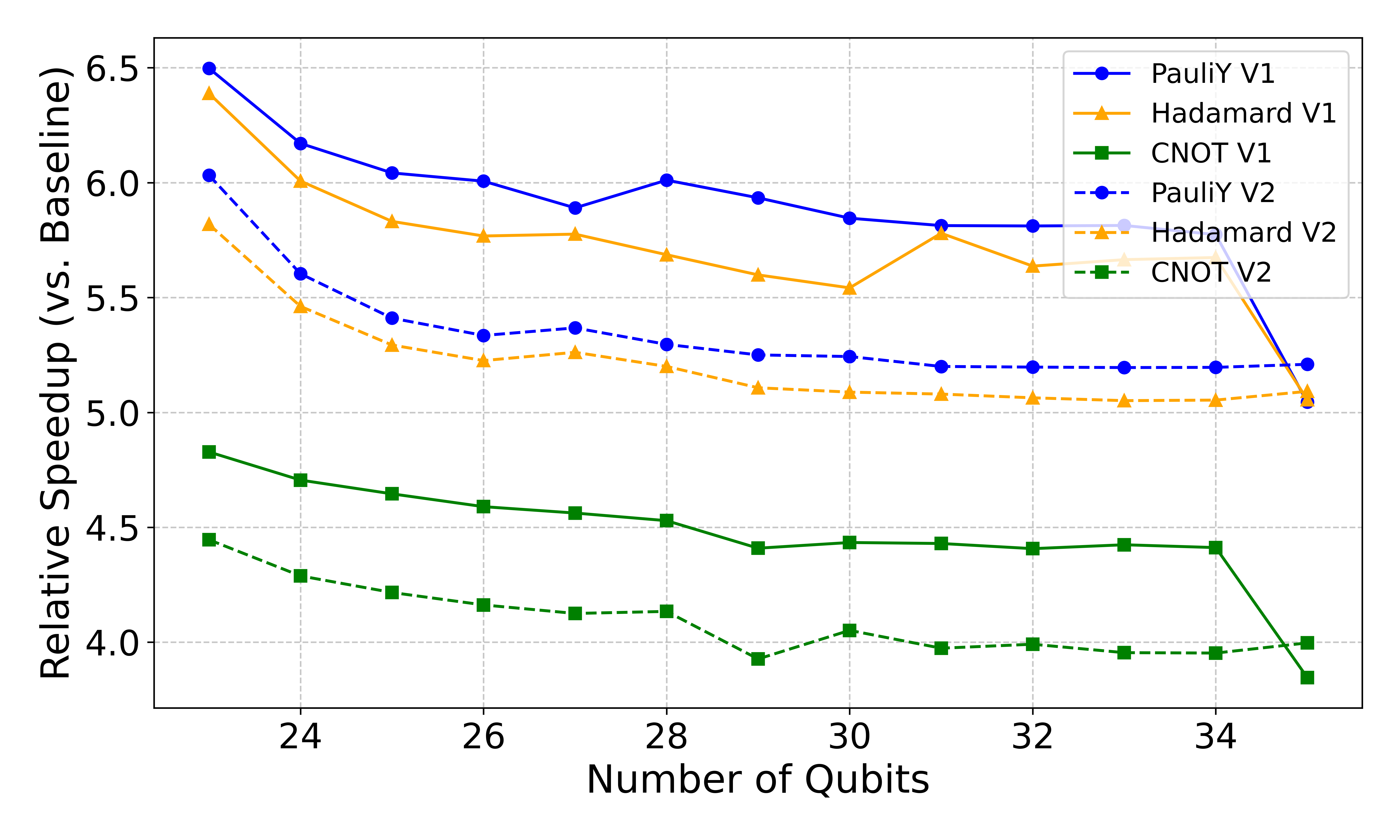}}
\caption{Relative speedup of the most optimized Hadamard, PauliY and CNOT gates vs. their respective QuEST baseline (default v3.7.0 build). V1, and V2 correspond to NUMA-aware local, and split allocations, respectively.}
\label{ben3}
\end{figure}

\subsection{Single- and Two-Qubit Operations} \label{singleQubit}
Figure \ref{ben1}(a) reports the absolute execution time per gate for a representative single-qubit gate (Hadamard). The baseline (default v3.7.0 build) QuEST results are compared against multiple optimization strategies, and the effectiveness of these optimizations is more evident in Figure \ref{ben1}(b), which shows the corresponding speedups relative to the baseline. The overall performance gains observed across our optimizations indicate the clear benefits of progressively more aggressive low-level tuning. Reordering the data into an AoS alone yields a modest speedup of $\sim$1.2$\times$ over the baseline. Adding AVX2 vectorization dramatically increases the speedup--$\sim$5$\times$ faster than baseline. When we switch from AVX2 to AVX-512 (with the AoS layout), the speedup improves further to $\sim$5.5$\times$. The best performance is achieved by the most aggressive optimization (AoS + AVX-512 + FMA + loop unrolling + prefetching), see Section~\ref{secTuning}, and Algorithm~\ref{alg:hadamard_full_ref}, reaching a speedup in the vicinity of 6$\times$ compared to the baseline. These gains reflect how each low-level improvement builds upon the previous ones and can be directly attributed to the underlying memory-access patterns of quantum operations. The restructured AoS format enhances spatial locality, which in turn enables efficient SIMD loading of data. 

For single‐qubit operations, the in‐place matrix‐vector multiply (depicted in Figure~\ref{unitary6q}) follows a relatively regular and predictable access pattern that is well‐matched to cache‐prefetching and advanced SIMD optimizations (AVX2 and especially AVX-512). There is minimal branching in these kernels; every qubit undergoes a uniform sequence of single‐qubit transformations in parallel, leading to dense, vectorizable memory loads and stores. As every amplitude is processed and there is minimal branching, the kernel processing each target qubit traverses (and implicitly permutes) the entire state vector in a systematic fashion that maximizes cache reuse and vectorization across CPU cores--resulting in significant performance gains.
Overall, our findings underscore that while both memory reorganization and vectorization contribute to improved performance, the combination of the new AoS with NUMA-aware tuning and instruction-level tricks is most effective in leveraging the regularity of the memory access pattern to yield the largest speedup when scaling simulations to higher qubit numbers.

Building on the optimizations introduced in Figure \ref{ben1}(b), Figure \ref{ben3} extends this analysis to compare the three gate types (Hadamard, Pauli-$Y$, and CNOT) using the fully optimized kernel under the two NUMA memory allocation strategies (see Section~\ref{sec3B}). V1 (node‐0-first) remains faster than V2 (split allocation) up to the penultimate qubit size. In this region, the memory footprint fits comfortably within a single NUMA node, so the overheads of remote accesses are minimal, and local allocation yields lower latency. However, once we reach the largest qubit size (where one node’s memory no longer suffices), V1 is forced to allocate data on the remote node as well, causing its performance to degrade. At this point, V2 (which evenly distributes memory across both nodes) achieves comparable or better results by balancing the load between nodes. 
Another key observation is that speed-up improvements differ notably among the three gate types. For Hadamard and Pauli-$Y$ operations, we achieve higher speedups (up to 5.5-6.5$\times$) over the baseline compared to CNOT (which peaks around 4.5$\times$). This behavior is closely tied to the memory‐access patterns depicted in Figures~\ref{control6q2}.
In contrast to the single-qubit gates, which involve uniform, independent rotations on all amplitudes (ideal for vectorized execution), CNOT involves controlled operations, where certain branches of the matrix‐vector multiply are skipped depending on the control qubit’s state--only the amplitudes corresponding to a control bit of 1 are updated. This conditional behavior introduces branches in the matrix‐vector multiply, as certain computational paths are effectively skipped.  This branching logic and the more irregular, conditional nature of the data manipulation reduces opportunities for continuous SIMD operations and hampers consistent cache reuse. Consequently, although CNOT still benefits from multi‐threading and low-level tuning strategies, it does not achieve speedups as high as the simpler single‐qubit operations.
\begin{figure}[htbp]
\centerline{\includegraphics[width=0.5\textwidth]{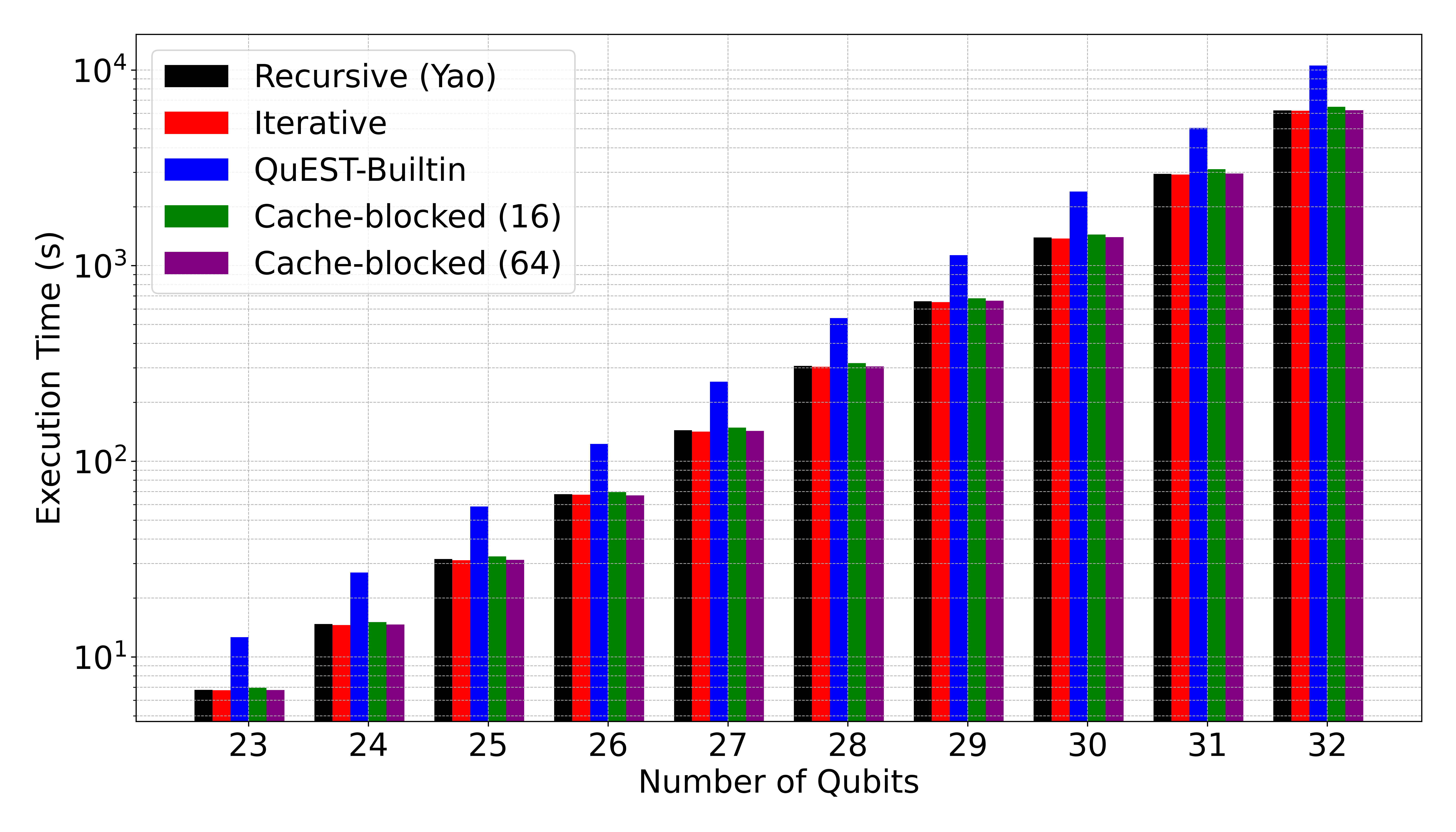}}
\caption{Execution time comparison for various baseline QFT implementations.}
\label{ben2}
\end{figure}

\subsection{Quantum Fourier Transform} \label{QFTSec}
To evaluate our approach on a more complex multi-qubit algorithm, we examined the QFT circuit. Figure~\ref{ben2} compares the execution times of the baseline QFT across ten qubit sizes using four different implementations: a divide-and-conquer recursive variant inspired by libraries like Yao/Qiskit, an iterative method, QuEST Built-in, and an optimized cache-blocked version with two blocking factors. The recursive implementation mirrors the mathematical definition of the QFT by recursively applying a Hadamard gate followed by controlled phase shifts. The iterative method, on the other hand, rearranges the qubit operations into nested loops instead of relying on recursion. The iterative layout can be more cache-friendly in certain scenarios due to its straightforward memory access pattern, representing differences of less than 0.5–1.5\% relative to the recursive approach across all qubit sizes. Although the built-in method is robust, it yields consistently higher execution times. At 23 qubits, the QuEST-Builtin runs $\sim$1.86$\times$ slower than the recursive/iterative variants, and at 32 qubits it is roughly 1.70$\times$ slower. This indicates that the hand-optimized approaches can provide a significant performance gain--roughly a 70–80\% reduction in runtime compared to the default library implementation. The QFT can be cache-blocked without extra gates by shifting its ending SWAP gates to the left, ensuring that all Hadamard gates operate on local qubits. Only the distributed SWAP gates (which are flipped vertically) incur communication overhead~\cite{Ref15}. In an effort to improve data locality, we implemented a cache-blocking technique with a blocking factor of 16, which is $\sim$3\% slower than the iterative approach at 23 qubits. At the upper end (32 qubits), the blocked variant is about 4.8\% slower than the iterative version. These results suggest that for the chosen block size, the overhead associated with managing cache blocks slightly outweighs the potential gains in memory locality, particularly as the data size grows. When we increased the block size to 64, performance remained virtually identical to the iterative baseline--evidence that the iterative method already exploits strong memory locality. The larger block size simply further amortizes data‐movement costs and maximizes cache utilization without adding significant overhead.

To evaluate the effectiveness of our optimizations, we compare two implementations of QFT--recursive and QuEST Built-in--as summarized in Table~\ref{tab:qft_speedups}. Overall, the recursive variant consistently achieves higher speedups ($\sim$1.8$\times$) than the Built-in ($\sim$1.7$\times$), indicating that its code structure may benefit more from compiler‐driven optimizations like inlining or function‐call elimination, giving it a minor advantage. Nevertheless, both implementations show robust improvements over their respective baselines due to low‐level enhancements. Hence, while both implementations benefit from our optimizations, the recursive QFT’s design appears to leverage these techniques slightly more effectively. Nonetheless, both QFT implementations are less amenable to extreme speedups compared to simple gate operations. The QFT’s series of dependent rotations and swaps impose a more sequential data access pattern, which limits the gains achievable from SIMD and caching techniques.
\begin{table}[htbp]
\centering
\caption{Speedup (baseline / optimized execution time) for recursive vs. QuEST's built-in QFT.}
\label{tab:qft_speedups}
\setlength{\tabcolsep}{10pt} 
\renewcommand{\arraystretch}{1} 
\begin{tabular}{@{}c c c@{}}
\toprule
Qubit & Recursive & Built-in \\
\midrule
23 & 1.79 & 1.67 \\
24 & 1.75 & 1.66 \\
25 & 1.76 & 1.65 \\
26 & 1.77 & 1.67 \\
27 & 1.78 & 1.67 \\
28 & 1.76 & 1.66 \\
29 & 1.77 & 1.68 \\
30 & 1.75 & 1.66 \\
31 & 1.76 & 1.65 \\
32 & 1.77 & 1.66 \\
\bottomrule
\end{tabular}
\end{table}
\begin{figure}[!htbp]
\centerline{\includegraphics[width=0.5\textwidth]{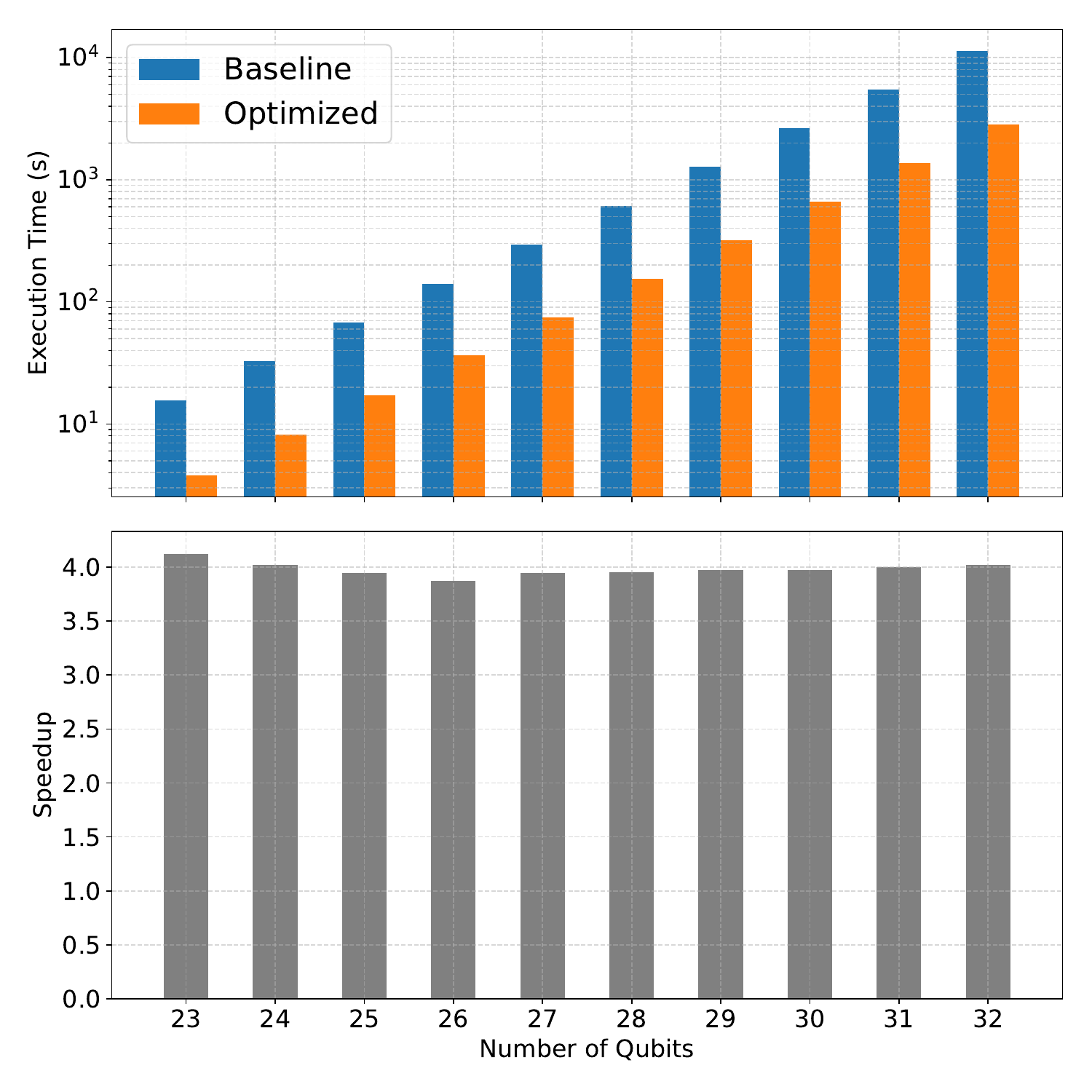}}
\caption{Comparison of execution times and speedup for baseline versus fully optimized RQC implementations.}
\label{rqc}
\end{figure}

\subsection{Random Quantum Circuits} \label{RQCSec}
Next, we implemented an RQC generator to stress-test the simulator. This setup allows us to quantify the impact of our low-level optimizations on both the dynamic generation of random single‑qubit gate sequences and the highly entangled execution of CNOTs connecting qubit $i$→$(i+1)\bmod n$. Here, each circuit comprises a fixed number of layers ($L=5$), and we sweep the register size $n$ from 23 to 32 qubits. We measured the performance of executing these circuits end-to-end, including state initialization, random gate generation, and state destruction, using the Google Benchmark framework for rigorous timing. Figure~\ref{rqc} presents the results for the baseline and optimized RQC executions. Despite the randomness of the circuits, their structure is fairly regular across qubits and layers, and our optimized simulator achieves a consistent speedup of $\sim$4$\times$ over the baseline. The RQC’s uniform layering and parallel gate structure are extremely well-suited to our low-level optimizations, which greatly reduce cache misses and overlap memory accesses with computation, thereby yielding a substantial boost in overall simulation speed. By contrast, as discussed above, the QFT’s inherently sequential structure and greater arithmetic complexity, which is primarily due to the interdependent controlled phase rotations introduces bottlenecks that constrain the effectiveness of SIMD strategies. Loop unrolling and prefetching do not capitalize on these operations as efficiently, and the sequential dependencies introduce communication overhead that further hinders performance improvements-limiting its speedup to $\sim$1.8$\times$. This comparison highlights that the effectiveness of our optimizations can depend on the circuit’s structure--highly parallel circuits like RQCs reap the most benefit, whereas circuits with serial dependencies see more moderate gains.

QFT is primarily compute-bound and particularly sensitive to vectorization, while RQC is memory-bound and strongly exposes data layout and NUMA effects. This complementary pairing has been used in prior simulation studies to isolate the effects of different architectural and kernel-level optimizations~\cite{queen2024,Ref12,Ref4}. However, these two workloads alone do not fully capture the behavior of structured algorithmic circuits. To broaden the evaluation scope, we additionally consider Grover and Shor-like periodicity estimation circuits, which introduce distinct entanglement and control patterns and offer a more representative view of end-to-end performance.

\begin{figure}[!htbp]
\centerline{\includegraphics[width=0.5\textwidth]{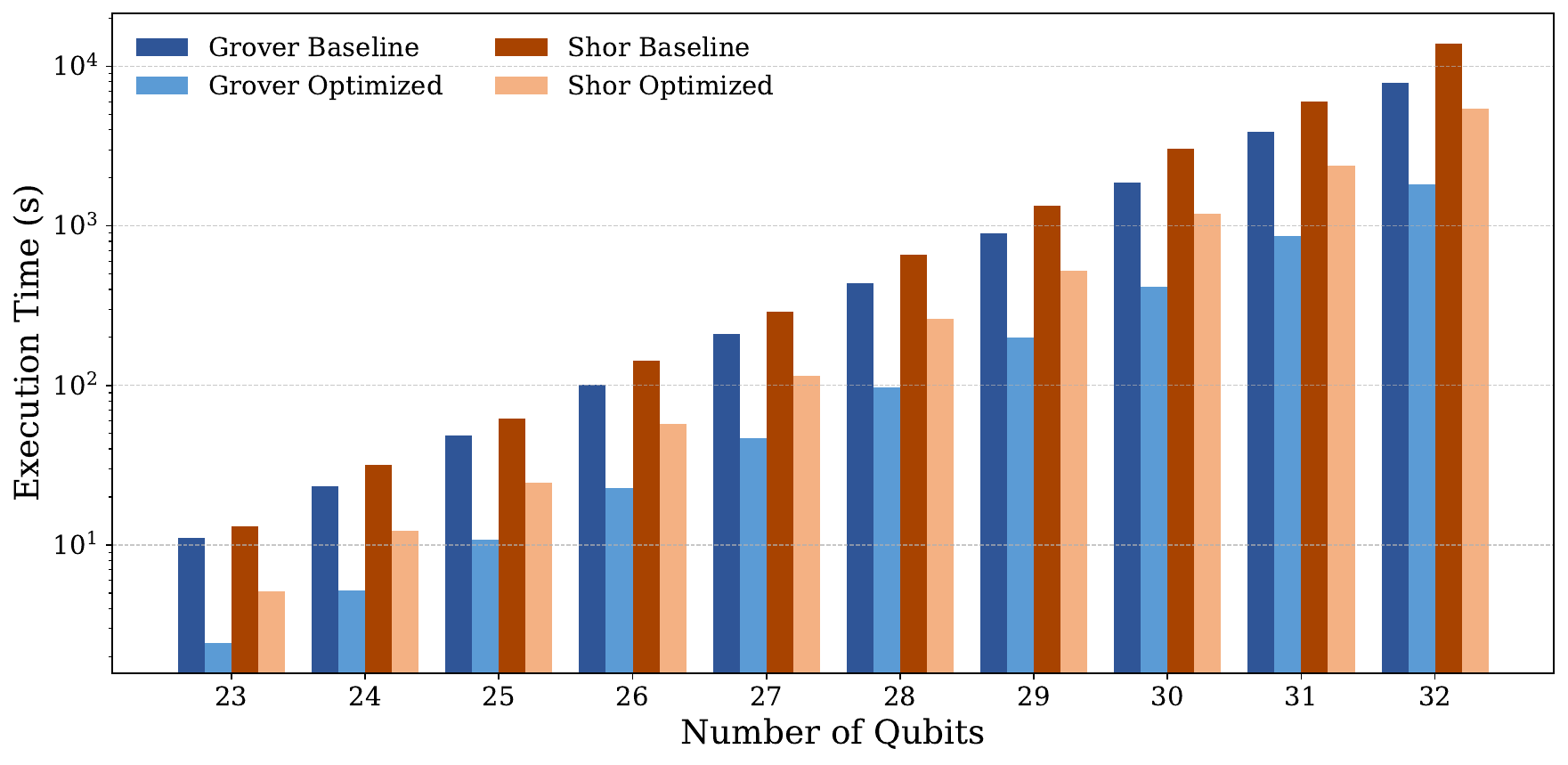}}
\caption{Comparison of execution times for baseline versus fully optimized Grover and Shor-like implementations.}
\label{GS}
\end{figure}
\subsection{Grover and Shor-like Workloads} \label{GSSec}
The execution time comparison presented in Figure~\ref{GS} and the corresponding speedup in Table~\ref{tab:speedup_grover_shor} clearly illustrate the impact of the optimized kernel on two representative circuit-level workloads, Grover and Shor. For both circuits, execution time increases exponentially with the number of qubits, as expected from the underlying $2^n$ state vector scaling. However, the optimized implementation achieves a substantial reduction in runtime across the entire range of qubit counts, reflecting the combined effect of NUMA-aware scheduling, AVX-512 vectorization, and cache-aware prefetching. The gains are particularly evident at higher qubit numbers, where memory bandwidth becomes the dominant bottleneck.

For Grover, which has relatively shallow depth and lower gate density, the optimized kernel delivers a stable speedup of approximately $4.3$–$4.6\times$ from 23 to 32 qubits. This indicates that the vectorized Hadamard, prefetching strategy, and NUMA-aware scheduling effectively sustain high memory throughput across all problem sizes. For the Shor-like circuit, which is substantially more gate-dense and involves $\mathcal{O}(n^2)$ controlled operations and an inverse QFT, the achieved speedups are lower but still significant, consistently around $2.5\times$ across the entire range. This reflects the heavier synchronization and memory traffic inherent in Shor, while still demonstrating a clear performance gain from the optimized kernels. These results confirm that the applied optimizations scale robustly for both low-depth and highly entangling algorithmic workloads.
\begin{table}[!htbp]
\centering
\caption{Speed-up factors (Baseline / optimized) for Grover and Shor-like algorithms.}
\label{tab:speedup_grover_shor}
\begin{tabular}{ccc}
\toprule
\textbf{Qubit} & \textbf{Grover Speed-up} & \textbf{Shor Speed-up} \\
\midrule
23 & 4.58$\times$ & 2.57$\times$ \\
24 & 4.50$\times$ & 2.62$\times$ \\
25 & 4.46$\times$ & 2.51$\times$ \\
26 & 4.46$\times$ & 2.51$\times$ \\
27 & 4.49$\times$ & 2.54$\times$ \\
28 & 4.50$\times$ & 2.52$\times$ \\
29 & 4.49$\times$ & 2.55$\times$ \\
30 & 4.48$\times$ & 2.54$\times$ \\
31 & 4.50$\times$ & 2.50$\times$ \\
32 & 4.34$\times$ & 2.56$\times$ \\
\bottomrule
\end{tabular}
\end{table}

\section{Discussion} \label{disc}

\textbf{\textit{Scaling.}}
Our experiments only include configurations with 2 NUMA nodes.
While this does not represent the maximum number of NUMA regions found in modern machines, we expect our evaluation to cover most practical configurations of them nonetheless, for two main reasons.
Firstly, when more compute power is needed, modern systems often reach for CPUs with more cores per socket, or accelerators such as GPUs or TPUs, rather than large socket counts per node.
Machines that need a large number of CPUs or GPUs usually fall directly under the distributed computing paradigm, where shared memory is no longer applicable~\cite{Ref21}.
Secondly, in the presence of several NUMA regions on 2 sockets, the penalties paid for going across sockets are much larger than when going across memory controllers on the same socket.
We observe for example on ARCHER2~\cite{Ref22}, a machine running dual-socket AMD\textregistered{} EPYC 7742 processors nodes with 8 NUMA regions (4 per processor), that going across sockets in the same node costs a 30\% decrease in memory bandwidth--similar to the performance
of the system we used for benchmarking--while going across NUMA regions on the same socket costs 2\% to 5\%.
This loss is so much smaller than going across sockets, that functionally this 8-region machine behaves like a 2-region configuration in practice.

\textbf{\textit{Portability.}}
Most elements of our optimisation stack are machine-agnostic. NUMA-aware page allocation--implemented with \texttt{mmap} + \texttt{mbind} -- works on any Linux host that ships \textit{libnuma}; when the library is missing or the system is single-socket, an \texttt{\#ifdef HAVE\_LIBNUMA} guard transparently falls back to plain \texttt{mmap}. Vectorisation is selected at compile-time: the fastest path uses AVX-512 intrinsics (\texttt{\_\_m512d}, \texttt{\_mm512\_*}) available on recent Intel processors (Skylake-X, Cascade Lake, Ice Lake, Sapphire Rapids). For wider coverage we supply: (i) an AVX2 variant built with \texttt{\_\_m256d}, (ii) a compiler-guided fallback that relies on \texttt{\#pragma omp simd} when no explicit SIMD ISA is present, and (iii) automatic ISA dispatch via GCC/Clang function multiversioning (\texttt{\_\_attribute\_\_((target("avx512f")))}) \emph{or} build-time flags such as \texttt{-mavx512f}, \texttt{-mavx2}, and \texttt{-march=native}, which lets the compiler pick the widest instruction set supported by the \emph{build host} automatically.
Thread scheduling stays portable because the loop is statically divided across cores, but the simple “first-half to node 0” binding must be replaced by a topology query on machines whose cores are not split 50/50. Finally, the two-cache-line prefetch distance is exposed as a tunable constant (\texttt{PREFETCH\_AHEAD\ 32}) so it can be retuned for processors with different LLC sizes or memory latencies. In short, every layer that drives the bulk of the speed-up either ports unchanged or degrades gracefully, enabling consistent acceleration on both modern Intel and AMD servers.

\section{Conclusion} \label{conc}
Our work bridges the gap between high-level algorithmic efficiency and low-level, architecture-specific performance enhancements in classical quantum simulation. We significantly extend the robust foundation of the QuEST simulator (v3.7.0) by developing a NUMA-aware, vectorized execution model explicitly optimized for modern multi-core processors. By restructuring the state vector to an AoS layout and applying a suite of targeted optimizations--including NUMA-aware memory allocation, explicit thread pinning, AVX-512 vectorization through intrinsic instructions, aggressive loop unrolling, and explicit memory prefetching--we achieved substantial and consistent speedups across a range of workloads.
For low-level primitives, we obtained speedups of $5.5$–$6.5\times$ for single-qubit gates and $4.5\times$ for two-qubit gates across 23–35 qubits. For circuit-level benchmarks, the optimized kernels delivered a $4\times$ speedup on RQC and approximately $1.8\times$ on QFT circuits. Extending beyond these benchmarks, we further evaluated two representative algorithmic workloads: Grover’s search and a Shor-like periodicity estimation circuit. These structured circuits, involving different entanglement and control patterns, exhibited stable performance gains of approximately $4.3$–$4.6\times$ for Grover and $2.5\times$ for Shor across 23–32 qubits, confirming the generality and robustness of the optimizations.
Our results compare favorably against related efforts such as HpQC and PAS. HpQC similarly employs AoS and AVX2 vectorization, reporting speedups of $2.20\times$ on QFT and $1.74\times$ on RQC, while PAS incorporates hybrid vectorization and fast bitwise arithmetic, reaching $8.69\times$ on QFT and $2.62\times$ on RQC. Unlike these approaches, our simulator integrates explicit NUMA-aware memory placement and thread affinity, substantially reducing inter-node latency and improving bandwidth utilization--features absent in both HpQC and PAS. 
Furthermore, while PAS and HpQC are not openly available, our implementation is fully open-sourced, promoting transparency, reproducibility, and community-driven improvements. The codebase is hosted in our group’s repository~\cite{Ref23} and will be integrated into mainstream QuEST following completion of compatibility and performance testing. This integration will enable a direct, quantitative comparison of our optimizations against the latest QuEST v4.0 release~\cite{Ref24}, which introduced new algorithmic features but not the low-level NUMA-aware and SIMD-based tuning presented here. 
In summary, this work provides a modular and extensible optimization framework that yields consistent performance gains across diverse workloads--from primitive gates and random circuits to structured algorithmic circuits such as Grover and Shor. These results highlight the importance of co-optimizing memory locality, vectorization, and thread placement in accelerating classical quantum simulations on modern multi-socket architectures.

\textbf{\textit{Future Work.}}
Going forward, we aim to maximize single-node performance and scalability through automation of vectorized kernel generation and compile-time-driven prefetching strategies, while extending support to larger shared-memory systems and heterogeneous accelerator-based platforms. In the near term, we plan to adapt NUMA-aware optimization concepts to GPUs, leveraging their shared emphasis on data locality. Although NUMA originated in CPU architectures, its core principles--memory locality and task affinity--are increasingly applicable to GPUs and can inform performance-aware execution across heterogeneous platforms~\cite{Ref25}.
Next, we will explore how emerging CXL memory expanders--i.e., CPU-less NUMA nodes--impact the performance and applicability of the techniques introduced in this work. While CXL clearly extends the memory capacity of a single node, its effect on simulation performance and the strategies required to optimize it remain open questions. Finally, we will explore low-level and locality-sensitive challenges in distributed multi-node, multi-GPU environments that leverage MPI and RDMA for communication.

\section*{Acknowledgment}
This work was supported by the EPSRC under grant EP/W032635/1 as part of the RoaRQ (Robust and Reliable Quantum) project. The authors thank the RoaRQ consortium for providing financial support and a collaborative environment that enabled the development of the methods and results presented in this paper. 



\vspace{12pt}

\bibliographystyle{IEEEtran}
\bibliography{IEEEabrv,mybib}

\end{document}